\documentclass[aps,pra,twocolumn,amsmath,amssymb,showpacs,superscriptaddress]{revtex4-1}
\usepackage{graphicx}
\usepackage{dcolumn}
\usepackage[mathlines]{lineno}
\usepackage{hyperref}
\usepackage{bm}
\usepackage{epstopdf}
\usepackage{epsfig}
\usepackage{bbold}

\newcommand{\bra}[1]{\ensuremath{\left\langle#1\right|}}
\newcommand{\ket}[1]{\ensuremath{\left|#1\right\rangle}}
\newcommand{\braket}[2]{\ensuremath{\left\langle #1 | #2 \right\rangle}}

\newcommand{\mean}[1]{\ensuremath{\left\langle#1\right\rangle}}
\newcommand{\TE}{\ensuremath{\ket{\mathrm{TE}_{\ell}}}}
\newcommand{\TM}{\ensuremath{\ket{\mathrm{TM}_{\ell}}}}
\newcommand{\HEe}{\ensuremath{\ket{\mathrm{HE}^e_{\ell}}}}
\newcommand{\HEo}{\ensuremath{\ket{\mathrm{HE}^o_{\ell}}}}

\usepackage{color}

\newcommand{\TK}[1]{\textcolor{black}{ #1}}

\newcommand{\BP}[1]{\textcolor{black}{ #1}}
\usepackage{color}
\newcommand{\tk}[1]{\textcolor{black}{ #1}}

\begin{document}

\title{Process tomography of quantum channels using classical light}

\author{Bienvenu~Ndagano}
\affiliation{School of Physics, University of the Witwatersrand, Private Bag 3, Wits 2050, South Africa} 
\author{Benjamin~Perez-Garcia}
\affiliation{School of Physics, University of the Witwatersrand, Private Bag 3, Wits 2050, South Africa}
\affiliation{Photonics and Mathematical Optics Group, Tecnol\'{o}gico de Monterrey, Monterrey 64849, Mexico}
\author{Filippus~S.~Roux}
\affiliation {School of Physics, University of the Witwatersrand, Private Bag 3, Wits 2050, South Africa}
\affiliation{CSIR National Laser Centre, P.O. Box 395, Pretoria 0001, South Africa} 
\author{Melanie~McLaren}
\affiliation {School of Physics, University of the Witwatersrand, Private Bag 3, Wits 2050, South Africa}
\author{Carmelo~Rosales-Guzman}
\affiliation {School of Physics, University of the Witwatersrand, Private Bag 3, Wits 2050, South Africa}
\author{Yingwen~Zhang}
\affiliation{CSIR National Laser Centre, P.O. Box 395, Pretoria 0001, South Africa} 
\author{Othmane~Mouane}
\affiliation {School of Physics, University of the Witwatersrand, Private Bag 3, Wits 2050, South Africa}
\author{Raul~I.~Hernandez-Aranda}
\affiliation{Photonics and Mathematical Optics Group, Tecnol\'{o}gico de Monterrey, Monterrey 64849, Mexico}
\author{Thomas~Konrad}
\affiliation{School of Chemistry and Physics, University of KwaZulu-Natal, Private Bag X54001, Durban 4000, South Africa} 
\author{Andrew~Forbes}
\email[Corresponding author: ]{andrew.forbes@wits.ac.za}
\affiliation{School of Physics, University of the Witwatersrand, Private Bag 3, Wits 2050, South Africa}

\date{\today}

\begin{abstract}
	\noindent \textbf{High-dimensional entanglement with spatial modes of light \TK{promises}  increased security and information capacity over quantum channels. 
	Unfortunately, entanglement decays due to perturbations, \TK{corrupting quantum links which cannot be repaired} without a tomography of the channel.  Paradoxically, the channel tomography itself is not possible without a working link.  Here we overcome this problem with a robust approach to characterising quantum channels by means of classical light. Using free-space communication in a turbulent atmosphere as an example, we show that the state evolution of classically entangled degrees of freedom is equivalent to that of quantum entangled photons, thus providing new physical insights into the notion of classical entanglement. \TK{The analysis of  quantum channels by means of classical light \TK{in real time} unravels stochastic dynamics in terms of pure state trajectories and thus enables precise} quantum error-correction in short and long haul optical communication, in both free-space and fibre.} 
	
\end{abstract}

\maketitle


\section{\label{sec:level1}Introduction}

Quantum correlations have become a ubiquitous resource in short and long-range communication using photons as carriers of quantum information (qubits). The most significant developments have been realised using polarisation as the  degree of freedom (DoF) of choice \cite{Ursin2007,Ma2012a,Yin2012,Herbst2015}; \TK{the two components of the polarisation vector of a photon are} robust against atmospheric perturbations, and can easily be controlled with wave-plates and polarising elements. Polarisation-based  quantum communication is, however,  limited to \TK{ a bandwidth of a single qubit per photon sent due to the low-dimensionality of polarisation}, and requires the sender and receiver to share a frame of reference.

\TK{Employing other degrees of freedom of light in} quantum protocols allows for more information to be packed onto single photons \cite{Cerf2002}. The use of spatial modes of light to realise high dimensions 
has seen many notable advances, with orbital angular momentum (OAM) being the preferred DoF \cite{Dada2011a, Romero2012, Fickler2014, Zhang2016}. OAM forms a convenient basis, is easy to measure with phase only holograms \cite{Forbes2016}, and is conserved down to the single photon level \cite{Mair2001}. However, it is worth noting that despite its potential, entanglement based on spatial modes  poses challenges in its implementation.  In both free-space and optical fibres, modal crosstalk and the concomitant decay of entanglement are the main challenges.

In free-space quantum channels, spatial modes are adversely affected by atmospheric turbulence \cite{Malik2012, Rodenburg12}, which reduces the probability of detecting photons \cite{paterson2005,Gopaul2007,Tyler2009}, while the induced scattering among spatial modes \cite{Chen2016,Neo2016} leads to a loss of entanglement in the final state measured in a given subspace \cite{ibrahim2013, Roux2015}.  To circumvent the deleterious effects of turbulence, as well as the need for a shared reference frame, hybrid OAM and polarisation qubit states have been put forward as possible carriers for more robust communication. These hybrid states are rotation invariant, and have been used to demonstrate alignment-free, robust quantum communication, where qubits are encoded \TK{in} the two DoFs that are entangled \cite{Souza2008, D'Ambrosio2012, Vallone2014, farias2015}. 

To date, channels with \TK{two-dimensional} quantum states have been demonstrated over 144 km with polarisation \cite{Ursin2007}, \TK{ and with hybrid OAM and polarisation states} over 210 m in a controlled environment to minimise turbulence \cite{Vallone2014}, \TK{as well as} recently over 3 km across Vienna \cite{Krenn2015}. 
Fibre channels with two dimensional entangled spatial modes languish at the centimetre scale \cite{Loffler2011, Kang2012}, and no study to date  has managed to report on the transport of high dimensional entanglement in any practical sense, in either free-space or fibre. To advance further requires characterisation schemes that allows one to gain information on the channel, predict the effects of perturbations, and implement error-correction in real-time.

Process tomography is an essential tool to obtain knowledge about the action of a channel in general, and its effects on the propagation of entangled states in particular \cite{Mohseni2008}. At the single photon level, this characterisation is difficult to do, especially with entangled states: one needs the quantum link to work before it can be characterised, but having it characterised would be immensely helpful in getting it to work. Thus the process tomography of quantum channels in which (entangled) spatial modes are used  remains topical but challenging.

Here we demonstrate a simple approach to characterise a quantum channel using classical light. We exploit the \TK{non-separability} property of vector beams, so-called classically entangled light \cite{Goyal2013,Guzman-Silva2015a,Karimi2015}, to show that 
\TK{the state evolution of two classically and two quantum entangled degrees of freedom is in one-to-one correspondence in case the channel for both systems acts only on a single DoF}.
This proves that beyond the mathematical resemblance to \TK{ its quantum counter part}, classical entanglement does hold physical significance. As an example, we demonstrate that the transport and decay \TK{ of the classical entanglement of  vector beams and the quantum entanglement of a photon pair} are identical in a channel perturbed by atmospheric turbulence. 
Moreover, \TK{we show that the one-to-one correspondence between one-sided channels and entangled states, the so called Choi-Jamiolkowski isomorphism \cite{Jiang2013}, is also valid for classical light fields with arbitrary degree of entanglement}. Thus, a full characterisation of quantum channels can be obtained via state tomography of classically entangled light beams. 
This new technique \TK{enables one, for example, to } determine the action of turbulence, and other channels, on pairs of spatial modes for quantum and classical states of light, and replaces the usual process tomography in both cases.
Finally, we demonstrate the applicability of the tools in a proof-of-principle communication experiment employing classically entangled states, showing that the characterisation of the channel allows for information recovery and robust data transfer. 


\section{\label{sec:level2} Results}
\begin{figure}[h]
	\centering
	\includegraphics[width = 0.48\textwidth, height = 0.23\textheight]{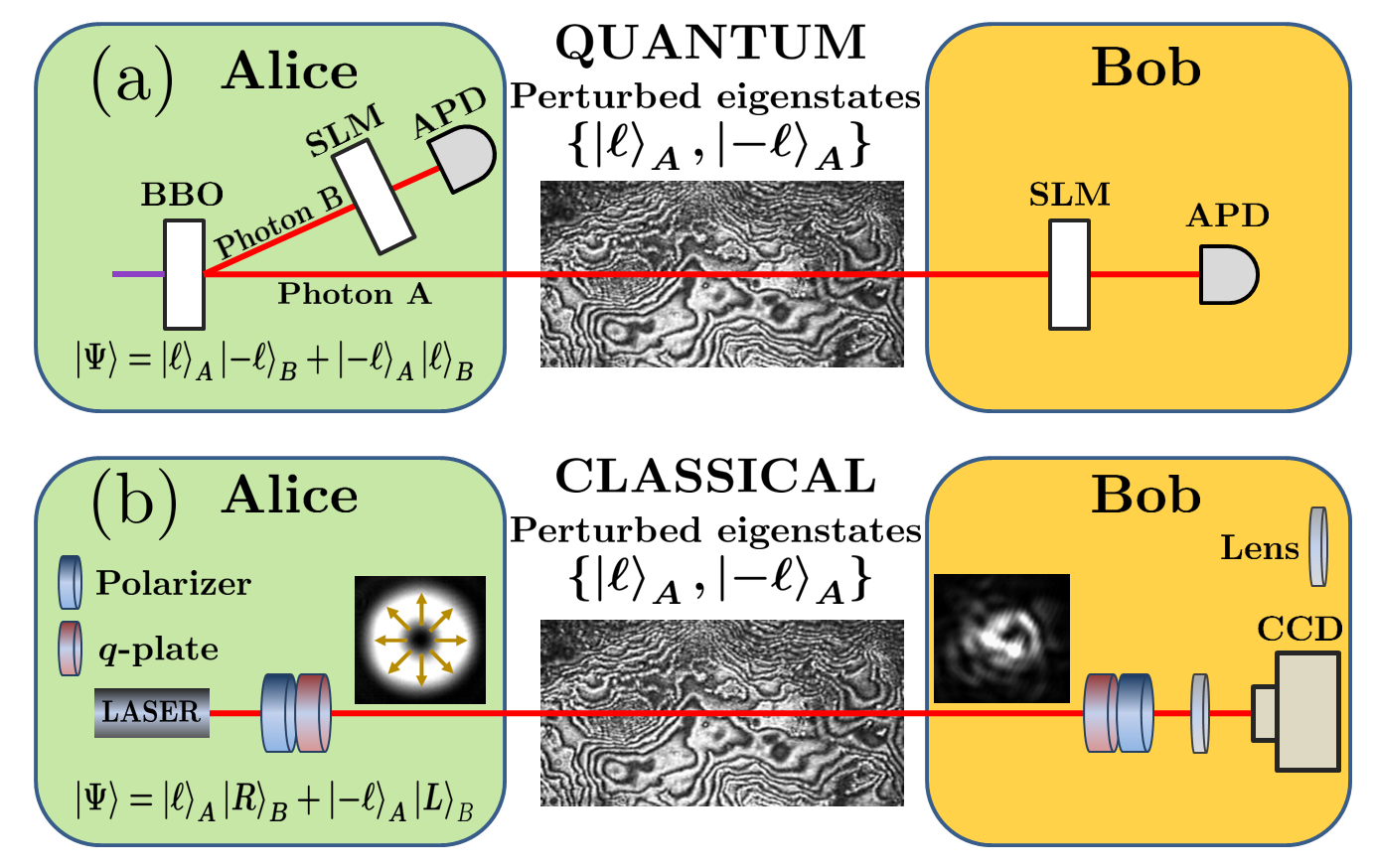}
	\caption{\textbf{Illustration of the concept.} (a) Alice generates two entangled photons using a SPDC process (see Methods), and sends one photon (Photon A) to Bob through a free-space turbulent channel, which affects the quantum correlations due to perturbations to photon A. (b) Equivalently, Alice sends information to Bob using a classically entangled bright light field (laser beam). The spatial degree of freedom (DoF A) is affected by the channel, while the polarisation (DoF B) is not. APD = avalanche photodiode, \BP{BBO = Beta Barium Borate (non-linear optical crystal)}, SLM = spatial light modulator and CCD = charge coupled device. }
	\label{Fig: Concept}
\end{figure}

\begin{figure*}[ht]
	\centering
	\includegraphics[width = 0.9\textwidth, height = 0.4\textheight]{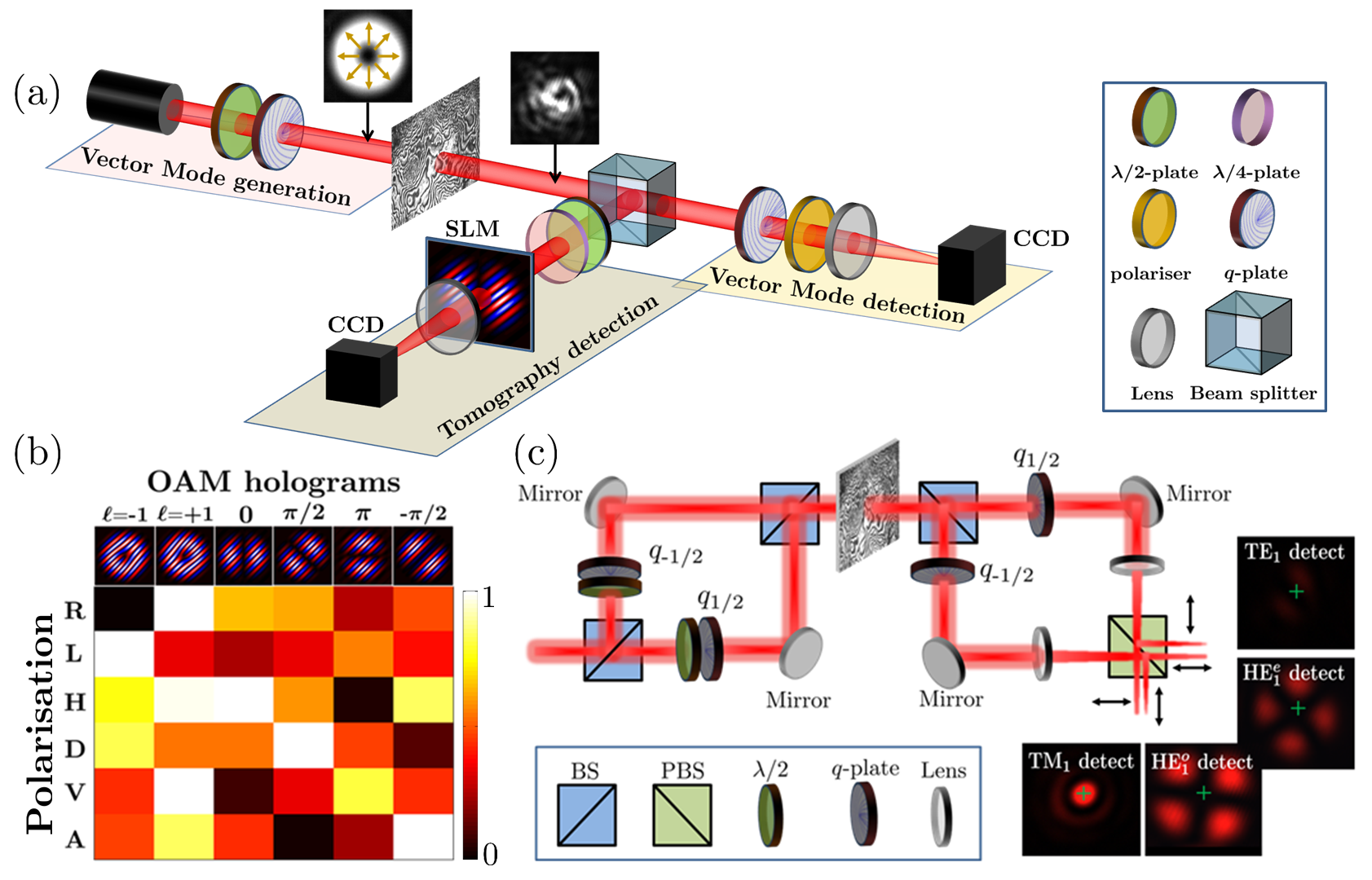}
	\caption{\textbf{Experimental setup.} (a) The entangled qubits were encoded using a $q$-plate and propagated through the turbulent conditions simulated with single Kolmogorov phase screens. The output state was analysed using a vector mode sorter that performs a decomposition into vector states. (b) Full state tomography of the perturbed state was performed by projecting polarisation states, selected with the quarter or half-wave plate, onto OAM states encoded on a spatial light modulator. The output of the projections were observed in the far-field using a camera. (c) Vector vortex modes are (de)multiplexed using two Mach-Zehnder interferometers, with the $q$-plates used to (de)encode hybrid qubits.}
	\label{Fig:setup}
\end{figure*}

\textbf{Concept}. Consider a typical scenario where quantum information is shared between two parties (Alice and Bob), as shown in Fig.~\ref{Fig: Concept}(a).  Alice generates two photons entangled in their spatial DoF, which we will take to be the OAM DoF.  Her bi-photon state can then be written as: $\ket{\Psi_{\ell}}_{{\mathrm{in}}} = \ket{\ell}_A \ket{-\ell}_B + \ket{-\ell}_A \ket{\ell}_B$.  She sends one photon to Bob, which passes through the channel; in this study we will consider a free-space link through a turbulent atmosphere as our example, but the concept is not restricted to this particular case and can be adapted to different channels.
It has been shown, theoretically and experimentally, that perturbations due to such a channel negatively affect the correlations between the photons, thereby decreasing the efficiency and security of the quantum communication link.  In this scenario, photon A experiences the channel while photon B does not.
The resulting state $\ket{\Psi_\ell}_{\mathrm{out}}$ of the photon pair  carries the complete information about the channel. This is due to the so-called Choi-Jamiolkowski isomorphism, and could be experimentally verified by teleporting states of single photons using the photon pair in state $\ket{\Psi_\ell}_{\mathrm{out}}$ as an entanglement resource. The resulting  teleportation channel would  reproduce the same state changes as the turbulence channel \cite{Dur2005}.

We claim that this quantum scenario has a classical equivalent, depicted in Fig.~\ref{Fig: Concept}.(b).  Here Alice prepares a classical beam that is non-separable in OAM and polarisation: $\ket{\Psi_{\ell}}_{{\mathrm{in}}} = \ket{\ell}_A \ket{R}_B + \ket{-\ell}_A \ket{L}_B$, sending the entire beam to Bob through the same channel.  In this formalism $A$ and $B$ now refer to the two degrees of freedom in the non-separable light field, and not to two photons in the entangled system.  But polarisation is not affected by turbulence, so the degree of freedom that experiences the deleterious effects of the channel is that of the spatial mode (A). In both cases only the states $\ket{\ell}_A$ and $\ket{-\ell}_A$ are affected.

The equivalence of the quantum and classical scenarios, together with the fact that the outgoing state in the quantum case contains the full information about the channel, strongly suggest that such non-separable states of light may be used to characterise the effect of the channel on the quantum state, an idea which we later validate theoretically and experimentally.\\

\textbf{Classically entangled states.} Our hybrid encoding space, described by the higher-order Poincar{\'e} sphere \cite{milione2011higher}, is formed from the tensor product of the infinite dimensional OAM and the two-dimensional polarisation Hilbert spaces. We are interested in states of the form

\begin{subequations}
	\begin{equation}
		\ket{\Psi_{\ell}}_{{\mathrm{in}}} = \alpha\ket{\ell}_A \ket{-\ell}_B + \beta\ket{-\ell}_A \ket{\ell}_B\ ,
		\label{eq: vector beam_quantum}
	\end{equation}
	\begin{equation}
	\ket{\Psi_{\ell}}_{{\mathrm{in}}} = \alpha\ket{\ell}_A \ket{R}_B + \beta\ket{-\ell}_A \ket{L}_B\ ,
	\label{eq: vector beam_classical}
	\end{equation}
\end{subequations}
where $|\alpha|^2 + |\beta|^2 = 1$. Equation \ref{eq: vector beam_quantum} defines a two-photon entanglement system expressed in the OAM basis, $\{\ket{-\ell},\ket{\ell}\}$, with each photon carrying  $\ell \hbar$ quanta of OAM \cite{Allen1992}. Equation \ref{eq: vector beam_classical} defines a vector vortex mode, which here will be the classical equivalent to the system defined in Eq.~\ref{eq: vector beam_quantum}. The basis states $\{\ket{L}, \ket{R}\}$ correspond the left and right circular polarisation states, respectively. 

 Among the many tools used to evaluate the degree of entanglement we choose the concurrence as our measure, as it has been shown to be effective in quantifying the degree of quantum and classical entanglement \cite{Wootters2001,McLaren2015}. For qubit pairs defined as in Eqs.~\ref{eq: vector beam_quantum} and \ref{eq: vector beam_classical}, this is given by:
\begin{equation}
	\mathcal{C}(\ket{\Psi_{\ell}}_{{\mathrm{in}}} ) = 2|\alpha\beta|\ .
	\label{conc1}
\end{equation}

\noindent\TK{\textbf{Channel tomography for turbulence}.} We consider an OAM state passing through turbulence \TK{ which causes modal dispersion and thus broadens}  the OAM spectrum. Here, \TK{for each instance of the stochastically varying turbulence}, the state evolution is a \TK{different} unitary transformation that maps pure states onto pure states \TK{and takes the form} (see Supplementary Information):
\begin{equation}
\ket{\ell} \rightarrow \sum_{\ell'} p_{\ell-\ell'} \ket{\ell'},
\end{equation}
where $p_{\ell-\ell'}$ are the modal weightings. Thus, a given input vector vortex mode $\ket{\Psi_\ell}_{\rm in}$ propagating through a turbulent channel will be transformed as follows:
\begin{eqnarray}
	\ket{\Psi_{\ell}}_{{\mathrm{in}}} \xrightarrow\  \alpha\sum_{\ell'}p_{\ell-\ell'}\ket{\ell'}\ket{R} + \beta\sum_{\ell'}p_{-\ell-\ell'}\ket{\ell'}\ket{L},
	\label{eq: state expansion}\\
	\begin{split} =  \alpha p_{0} \ket{\ell}\ket{R} &+ \alpha p_{2\ell} \ket{-\ell}\ket{R} + \beta p_{-2\ell} \ket{\ell}\ket{L}\\
		&+ \beta p_{0} \ket{-\ell}\ket{L} + \sum{(\ldots)}\ .\end{split}
\end{eqnarray}
Note that we omit the subscripts A and B for simplicity of notation.  \TK{Filtering for OAM values $-\ell$ and $\ell$ yields the unnormalised final state} 
\TK{\begin{equation}
	\begin{split}
		\ket{\Psi_{\ell}}_{{\mathrm{out}}}  = \alpha p_{0} \ket{\ell}\ket{R} + \beta p_{-2\ell} \ket{\ell}\ket{L} &+ \alpha p_{2\ell} \ket{-\ell}\ket{R} \\
		& + \beta p_{0} \ket{-\ell}\ket{L}.
	\end{split}
	\label{Eq: outstateturb}
\end{equation}
}
\TK{ We can now read off the operator $\mathrm{M}$ that describes the action $\ket{\Psi_{\ell}}_{{\mathrm{in}}}\rightarrow\ket{\Psi_{\ell}}_{{\mathrm{out}}}=\mathrm{M}\otimes \mathbb{1} \ket{\Psi_{\ell}}_{{\mathrm{in}}}$ of the channel on the OAM degree of freedom:}
\TK{\begin{equation}
	\begin{split}
		\mathrm{M} = p_{0} \ket{\ell}\bra{\ell} + p_{-2\ell} \ket{\ell}\bra{-\ell} &+ p_{2\ell} \ket{-\ell}\bra{\ell} \\
		& + p_{0} \ket{-\ell}\bra{-\ell}.
	\end{split}
	\label{Eq: channelmat}
\end{equation}}
\TK{The one-to-one correspondence between the state (Eq. \ref{Eq: outstateturb}) of the vector beam  and the Kraus operator (Eq. \ref{Eq: channelmat}) for the OAM channel established here for classical light fields, is a manifestation of the Choi-Jamiolkowski isomorphism  known to also connect the quantum OAM channel (same operator M)  and the state of the qubit pair obtained from Eq. \ref{eq: vector beam_quantum}. }  It follows that the matrix elements of the \TK{quantum channel operator}, M, typically determined by process tomography \TK{using many photons, can now be monitored} by a state tomography of the \TK{classical} output state $\ket{\Psi_{\ell}}_{{\mathrm{out}}}$ \TK{in real time}. \\

\begin{figure}[t]
	\centering
	\includegraphics[width = 0.45\textwidth, height = 0.45\textheight]{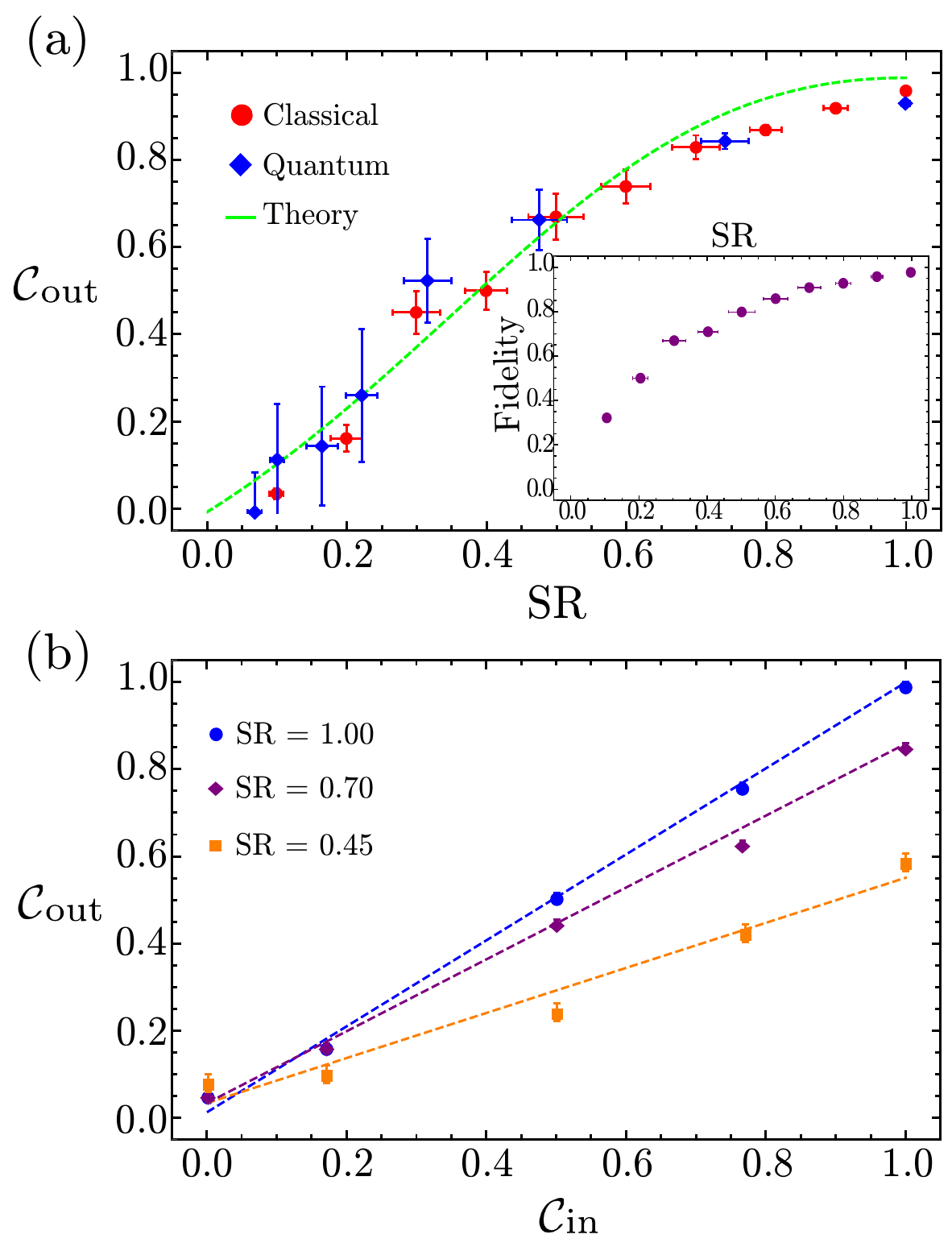}
	\caption{\textbf{Evolution of classical entanglement in turbulence.} (a) The measured concurrence variation after propagation through turbulence screens of various SR strengths (increasing from high to low SR values) for the classical (red markers) and quantum (blue markers) states, correctly follow the theoretical prediction for one of two entangled photons going through a turbulent channel (green dashed line). The fluctuations in SR arise from the statistical averaging of the turbulence screen for an encoded SR value. The inset shows the decay in fidelity of the output state with respect to a maximally entangled state, as a function of the turbulence strength. (b) The measured concurrence obeys the predicted linear mapping to the concurrence of the input state. Here we have shown this for three channels with SR = 1.00, 0.70 and 0.45.}
	\label{Fig: Evo Entangl}
\end{figure}

\textbf{Decay of classical entanglement in turbulence.} 
The entanglement of the  final state (Eq. \ref{Eq: outstateturb}) is given by the concurrence $\mathcal{C}(\ket{\Psi_{\ell}}_{{\mathrm{out}}} ) = 2|\alpha \beta||p_0^2 - p_{2\ell}p_{-2\ell}|/p$, which, can be expressed in terms of the \TK{entanglement} of the input state \TK{by}
\begin{equation}
	\mathcal{C}(\ket{\Psi_{\ell}}_{{\mathrm{out}}} ) = \mathcal{C}_{\mathrm{ch}}~\mathcal{C}(\ket{\Psi_\ell}_{\text{in}})\ ,
	\label{Eq: Concinturb}
\end{equation}
where \TK{$0\le\mathcal{C}_{\mathrm{ch}}=|p_0^2 - p_{2\ell}p_{-2\ell}|/p\le 1$    equals the percentage of entanglement preserved by the channel and is given by the output concurrence obtained from an initially maximally entangled state with probability $p= (2|p_0|^2 + |p_{2\ell}|^2 + |p_{-2\ell}|^2)/2$. }
For example, in weak turbulence where \TK{$|p_{2\ell}p_{-2\ell}|/p\ll |p_{0}^2|/p \approx 1$}, Eq. \ref{Eq: Concinturb} reduces to: 
\begin{equation}
	\mathcal{C}(\ket{\Psi_{\ell}}_{{\mathrm{out}}} ) \approx \mathcal{C}(\ket{\Psi_\ell}_{\text{in}})\ .
	\label{eq. conc_out/in}
\end{equation}
The initial and final entanglement are approximately the same, i.e., the photons remain entangled to each other. Conversely, in relatively strong turbulence where 
$|p_{2\ell}p_{-2\ell}|\approx|p_{0}^2|$, the concurrence ${\cal C}(\ket{\Psi_{\ell}}_{{\mathrm{out}}} )$ vanishes. 
The relation in Eq. \ref{Eq: Concinturb} has been derived in \cite{Konrad2007} for entangled photons.

The broadening of the OAM spectrum described by Eq. \ref{eq: state expansion} leads to inter-modal coupling among vector modes, resulting in a loss of entanglement with increasingly strong perturbations. Within the subspace of vector modes with $|\ell| = 1$, this coupling can be analysed using the four vector modes in this space. These modes are orthogonal and constitute a basis, analogous to the Bell basis, that can be used to encode information \cite{Milione2015e}. In optical waveguide theory, these modes are widely known as optical fibre modes \cite{Snyder1984}, which we label as:
\begin{eqnarray}
	\ket{\mathrm{TM}}_{1} &=& \frac{1}{\sqrt{2}}(\ket{1}\ket{R} + \ket{-1}\ket{L})\ , \label{Psi states1} \label{maxent} \\
	\ket{\mathrm{TE}}_{1} &=& \frac{1}{\sqrt{2}}(\ket{1}\ket{R} - \ket{-1}\ket{L})\ , \label{Psi states2}\\
	\ket{\mathrm{HE^{e}}}_{1} &=& \frac{1}{\sqrt{2}}(\ket{1}\ket{L} + \ket{-1}\ket{R})\ , \label{Phi states1}\\
	\ket{\mathrm{HE^{o}}}_{1} &=& \frac{1}{\sqrt{2}}(\ket{1}\ket{L} - \ket{-1}\ket{R})\ . \label{Phi states2}
\end{eqnarray}
By way of example, consider a $\ket{\mathrm{TM}}_{1}$ propagating in a strong turbulence regime. In the special case where \TK{$p_0 =~p_{2\ell} =~p_{-2\ell}$} (strong coupling), the final state reduces to
\TK{\begin{align}
	\ket{\Psi_1}_{\text{out}} &= \frac{1}{\sqrt{2}}p_0\left[\ket{1}\ket{R} + \ket{-1}\ket{L} + \ket{1}\ket{L} + \ket{-1}\ket{R}\right],\\
	&= \frac{1}{\sqrt{2}}p_0\left[\ket{\mathrm{TM}}_{1} + \ket{\mathrm{HE^{e}}}_{1} \right]\ ,\\
	&= \frac{1}{\sqrt{2}}p_0\left(\ket{1}+\ket{-1}\right)\otimes\left(\ket{R}+\ket{L}\right)\ ,
\end{align}}
which is separable (not entangled) i.e., the spatial and polarisation DoFs can be factorised. Equivalently in the quantum case, perturbations incurred by one of the two photons (modal dispersion and projection onto a subspace) will transform an initial entangled state into a final factorisable (separable) state.\\


\begin{figure*}[t]
	\centering
	\includegraphics[width = 0.9\textwidth, height = 0.43\textheight]{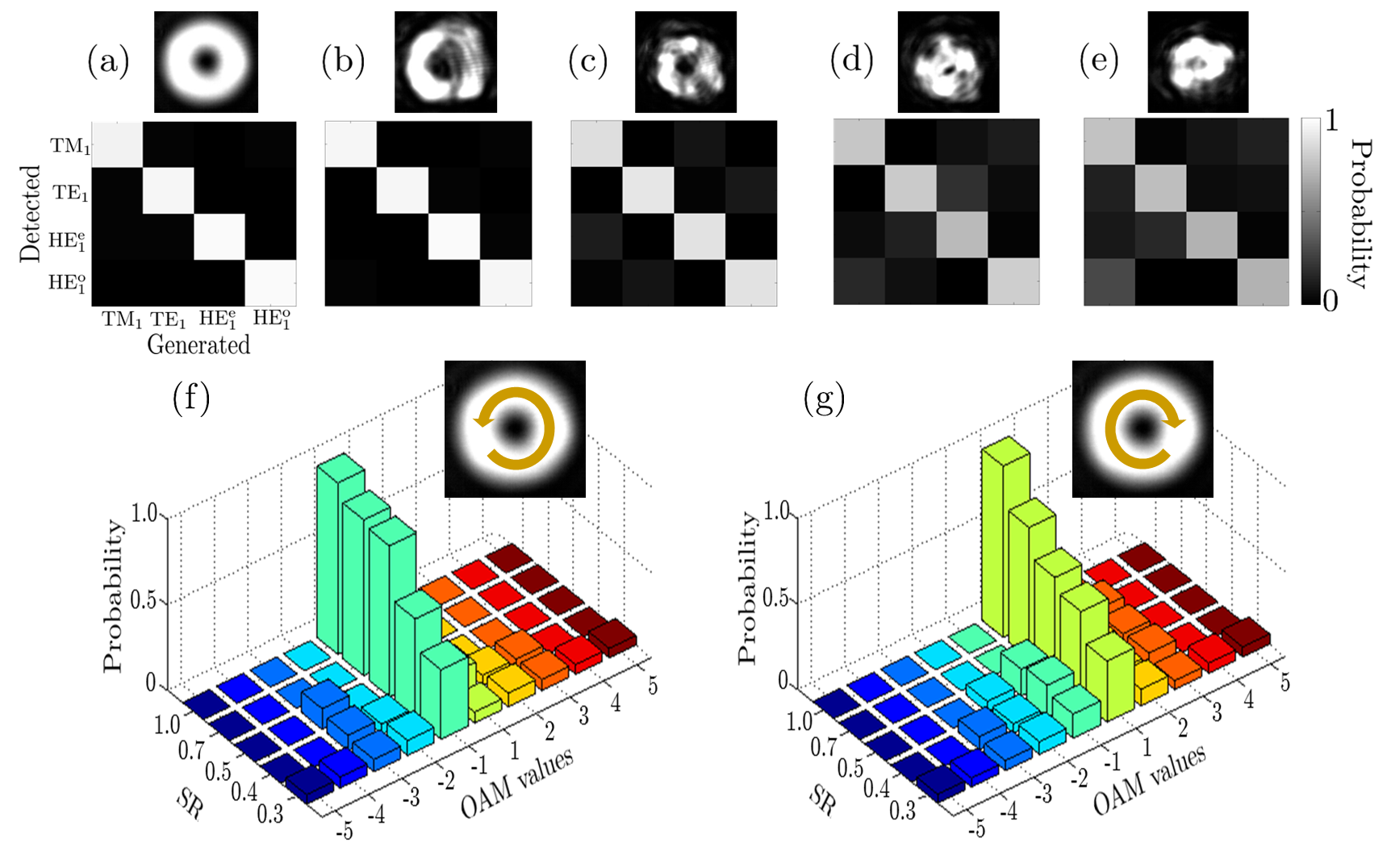}
	\caption{\textbf{Evolution of the mode spectrum in turbulence}. (a)-(e) Show vector mode dispersion for the four vector vortex modes of the $|\ell| = 1$ subspace for average SR values of \BP{1.0}, 0.7, 0.5, 0.4 and 0.3 respectively. The matrices show the evolution of the power distribution among the four vector detector for a given input \BP{mode}. (f) and (g) Show the effect of the turbulence on the spatial DoF, measured by modal decomposition of the $\ket{-1}\ket{L}$ and $\ket{1}\ket{R}$ input qubit states, respectively. The scattering of the OAM eigenstates is observed through a redistribution of energy between OAM modes with increasing SR.}
	\label{Fig: Evo Crosstalk}
\end{figure*}

\textbf{Classical and quantum experiments.} Here we demonstrate experimentally the equivalence between the evolution of classical and quantum entanglement in turbulence. Our classical experimental setup is illustrated in Fig.~\ref{Fig:setup} and comprises creation, propagation and detection stages.  In the creation step, the vector vortex mode is prepared either directly from a ``spiral laser'' \cite{Naidoo2016}, or by using wave plates and $q$-plates \cite{Marrucci2006} to transform a linearly polarised Gaussian beam into a vector vortex mode. We passed our vector mode through a turbulent channel (turbulence phase screen) that was made to vary with time, and analysed the output beam with two detection systems: a vector mode sorter to uniquely detect each of the maximally entangled modes, thus evaluating the amount of inter-modal coupling, and a tomography detector \cite{McLaren2015} to evaluate the concurrence \tk{(see Methods)}. In the quantum experiment, we used spontaneous parametric down conversion (SPDC) to produce two photons entangled in OAM, of which one was sent through a turbulent channel (See Methods). A state tomography of the two photons was performed to determine the evolution of the concurrence as a function of the turbulence strength. To account for fluctuations in the number of photons, we used an over-complete set of measurements to reconstruct the density matrix (see Methods).

In Fig. \ref{Fig: Evo Entangl}(a) we show the measured dependence of the concurrence of our quantum and classical states as a function of the degree of turbulence in the channel, together with the theoretical prediction from \tk{Eq.~\ref{concsr}}. We used a $\ket{\mathrm{TM}}_{1}$ vector mode and an $|\ell|=1$ maximally entangled OAM state as our equivalent classical and quantum systems, respectively. The experimental results for both the classical and quantum cases are in excellent agreement with the theory. Hence, the agreement between the classical and quantum experiments validates the equivalence of the quantum and classical models depicted in Fig.~\ref{Fig: Concept}. The inset in Fig. \ref{Fig: Evo Entangl}(a) shows the variation in the measured fidelity of a $\ket{\mathrm{TM}}_{1}$ vector mode in turbulence, computed with respect to a maximally entangled state. Furthermore, by varying the concurrence of the input vector mode in Fig.~\ref{Fig: Evo Entangl}(b), we experimentally confirmed for the first time, the existence of the Choi-Jamiolkowski isomorphism for spatial modes as summarised in Eq.~\ref{Eq: Concinturb}, i.e., that there is a linear relationship between $	\mathcal{C}(\ket{\Psi_{\ell}}_{{\mathrm{out}}} )$ and $	\mathcal{C}(\ket{\Psi_{\ell}}_{{\mathrm{in}}} )$.

The observed decay of entanglement with increasing turbulence, as predicted by Eq.~\ref{concsr}, is explained by examining the effects of turbulence on the mode spectrum: for example, in the classical case, with increasing turbulence strength, the scattering among vector vortex modes is increased, as seen in Figs.~\ref{Fig: Evo Crosstalk}~(a)-(e). This \tk{is due} to the spreading of the OAM spectrum,  as shown in Figs.~\ref{Fig: Evo Crosstalk}(f) and \ref{Fig: Evo Crosstalk}(g), which leads to crosstalk among vector vortex modes, as described by Eq.~\ref{Eq: outstateturb}, and similarly for the quantum \tk{case.}
\begin{figure}
	\centering
	\includegraphics[width = 0.48\textwidth, height = 0.35\textheight]{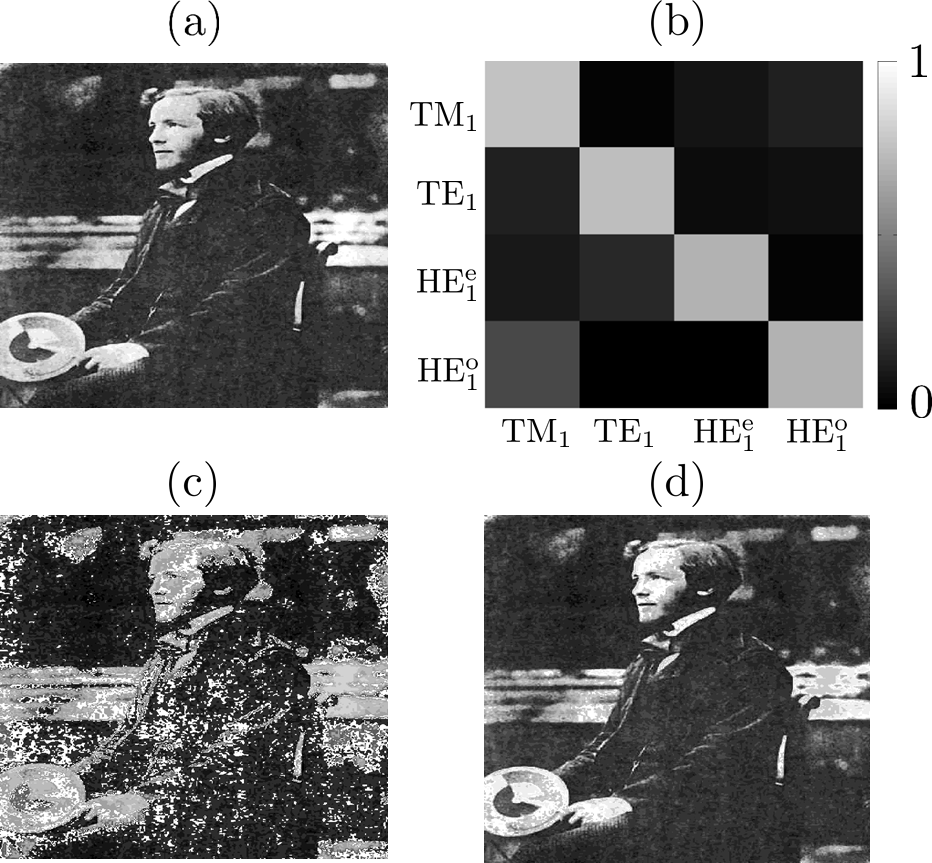}
	\caption{\textbf{Experiemental results for data transmission over turbulent channel.} (a) Using a four-bit encoding technique with four multiplexed vector modes, a 425$\times$513 pixels image of Maxwell was transmitted through a turbulent channel with $\mean{\mathrm{SR}} = 0.3$, and demultiplexed at the receiver's end. A threshold at 15\% of the maximum received signal was applied to the data to filter noise signals. By characterising the channel with a vector mode sorter, we obtained (b) the channel matrix. (c) The channel perturbation resulted in a distorted image at the receiver's end with a correlation coefficient of 64.2\%. (d) Correcting the image with the inverse channel matrix increased the correlation coefficient to 98.9\%.}
	\label{Fig: maxwell}
\end{figure}

Having characterised the impact of the channel on the states to be propagated, we used the setup to show a proof-of-principle data transmission over the channel. It has been recently shown that the non-separability of vector vortex \tk{beams} can be used to encode two bits of information simultaneously on the entangled DoFs \cite{Milione2015e}. In our scheme, we (de)multiplexed the four maximally entangled vector modes as shown in Fig. \ref{Fig:setup}(c). This allowed us to perform a four-bit encoding scheme based on these states. The encoded image in Fig. \ref{Fig: maxwell}(a) was transmitted through a turbulent channel with an average turbulence strength $\mean{\mathrm{SR}} = 0.3$.  Without any correction, the received image shows significant amounts of distortion, resulting in 
a 64.2\% correlation coefficient with respect to the encoded image [Fig. \ref{Fig: maxwell}(c)]. This is due to the intermodal coupling corrupting the encoded bits sequences, and resulting in state errors measured at the receiver's end. A practical advantage of studying the decoherence induced by a channel is the ability to mitigate perturbations through pre- or post-processing of the data. After propagation through a perturbing medium, the input and output states can be related by
\begin{equation}
\ket{\Psi_\ell}_{\mathrm{out}} = \text{M}\otimes\mathbb{1}\ket{\Psi_\ell}_{\mathrm{in}},
\end{equation}
where M is the channel operator of the system that contains all the information about the crosstalk induced by the medium. This matrix  \tk{ $\text{M}\otimes\mathbb{1}$} is graphically represented in Fig. \ref{Fig: maxwell}(b). Thus the perturbation can be cancelled by correcting the final state, $\ket{\Psi_\ell}_{\mathrm{out}}$, with \tk{M$^{-1}\otimes\mathbb{1}$ (see Supplementary Information)} . Using this correction technique, we obtained an image with an increased correlation coefficient of 98.9\%, as shown in Fig. \ref{Fig: maxwell}(d).

\section{\label{sec: level3} Discussion}
The characterisation of quantum channels is a \textit{sine qua non} to the implementation of practical quantum communication protocols. Perturbations from the environment constitute a hindrance in realising quantum links, particularly when using entanglement as a resource. Here we have described, as an example, the effects of atmospheric turbulence on entangled spatial modes. Using a classically equivalent system, we showed that the quantum channel can be characterised with bright classical sources.

Using vector vortex modes, so-called classically entangled light, we have proved that the state evolution of two entangled DoFs is identical to that of two photons entangled in one DoF, when propagating through atmospheric turbulence. As a corollary, in both the quantum and classical pictures, our models show an identical decay in entanglement correlation with increasing perturbations. This provides new insights into the notion of entanglement at the classical level; that is, beyond the mathematical non-separability of the DoFs, nature cannot \BP{distinguish} between classical and quantum entanglement in as far as characterising the channel is concerned. Furthermore, our work represents, to the best of our knowledge, the first side by side comparison of classical and quantum entanglement. We have shown this for atmospheric turbulence, but the approach can easily be generalised to other 
perturbations.

The decay of entanglement we observed in our results show that vector vortex modes are not resilient to atmospheric turbulence. This is further supported by the decay in fidelity of the final state after turbulence, measured with respect to a maximally entangled state (see Methods), as shown in the inset of Fig. \ref{Fig: Evo Entangl}(a). The decay of entanglement and fidelity we found is not in contradiction with a method to recover qubits encoded in two rotationally symmetric vector states \cite{D'Ambrosio2012} that was tested against the influence of turbulence \cite{farias2015}. This method uses a filter (post-selection) to eliminate all spatial crosstalk components generated by weak turbulence, resulting in the loss of photons. However, this post-selection approach does not provide a measure of resilience to turbulence as all modes can be recovered using this technique. 

We confirmed that the channel's impact on the quantum state can be determined from a single measurement of the maximally entangled state. Unlike in quantum optics experiments with entangled bi-photons in their spatial modes, here the degree of entanglement (non-separability) of our classical light may easily be adjusted with simple polarisation optics. This allows a sender (receiver) at the input (output) to predict the loss in correlation as a result of perturbation induced by the channel o quantum states with arbitrary degree of entanglement. This is a consequence of the Choi-Jamiolkowski isomorphism which, for the first time, has been demonstrated with spatial modes. This may pave the way for forward-error-correction and identification of an eavesdropper in quantum key distribution protocols through a noisy channel.

Using the tools we presented to characterise the channel, we demonstrated a simple prepare-and-measure protocol whereby data was encoded on classically entangled states, sent through the turbulent channel, decoded and corrected using the channel matrix.
Although this demonstration serves as a proof-of-principle, the techniques presented in this work can be applied to quantum error-correction. The characterisation of quantum channels through process tomography requires multiple measurements to be performed on the state over extended lengths of time - this is what gives rise to mixed states despite the unitary behaviour of the state evolution. 
\TK{The ability to detect the specific realisations of stochastic perturbations classically, enables an unraveling of the otherwise mixed-state dynamics in terms of pure states and allows unitary unraveling for perfect control.}
Using classical light, the process tomography measurements can then be done simultaneously (since there are many photons), allowing for real-time error-correction. 


In conclusion, we have proposed and demonstrated a classical approach to study the transport of a quantum entangled system in a perturbing channel.  Using free-space communication in a turbulent atmosphere as an example, we claimed and proved the equivalence of classical and quantum entanglement when characterising a channel. This equivalence was demonstrated in a direct comparison of the decay in the degree of entanglement for  a bright classical vector beam and entangled photons. In this paradigm, we showed that the process tomography of quantum channels, requiring multiple measurements on the quantum state, can be replaced by a state tomography on the classical beam. This process tomography of a communication channel using classical light can be done in real-time, and implemented in quantum links for real-time error-correction. Lastly, we have proved, again using classical light, the Choi-Jamiolkowski isomorphism: given a channel, the decay in entanglement of a quantum state can be predicted from that of a maximally entangled state,  through a linear relationship. 

Through our theoretical analysis and experimental investigations, we have proved that classical entanglement is more than a mathematical non-separability; it has physical properties which, in some cases, nature itself cannot differentiate from those of its quantum counterpart. 


\section{\label{sec: level4}Methods}
\textbf{Generation and detection of vector vortex beams using a $q$-plate}. The generation of vector vortex beams has been made convenient with the invention of $q$-plates. These are phase plates with locally varying birefringence that gives rise to a coupling between SAM and OAM through the Pacharatnam-Berry geometric phase \cite{Marrucci2006}. The encoding of entangled qubits with a $q$-plate is summarised by the following transformation rules: 
\begin{eqnarray}
	\left| \ell, L \right\rangle \xrightarrow{q\text{-plate}} \left| \ell + 2q, R \right\rangle, \label{eq:Qplate1}\\
	\left| \ell, R \right\rangle \xrightarrow{q\text{-plate}} \left| \ell - 2q, L \right\rangle,
	\label{eq:Qplate2}
\end{eqnarray} 
where $q$ is the topological charge of the $q$-plate. The four vector vortex modes of a given $|\ell|$ subspace are non-separable superpositions of qubit states generated as in (\ref{eq:Qplate1}) and (\ref{eq:Qplate2}) with the $\ket{L}$ and $\ket{R}$ input components phase shifted by $0$ or $\pi$. By transforming a linearly polarised Gaussian beam, the \TE\ and \TM\ are generated with a $q$-plate with $+|q|$ topological charge, while the \HEe\ and \HEo\ are \BP{generated} with one having $-|q|$ topological charge.\\
In addition to their encoding function, $q$-plates can also be used as decoders. This is achieved by simply reversing the generation process outlined in (\ref{eq:Qplate1}) and (\ref{eq:Qplate2})
\begin{eqnarray}
	\ket{\ell + 2q, R}\xrightarrow{q\text{-plate}} \ket{\ell + 2(q-q'), L}, \label{eq:Qplate1b}\\
	\ket{\ell - 2q, L}\xrightarrow{q\text{-plate}} \ket{\ell - 2(q-q'), R}.
	\label{eq:Qplate2b}
\end{eqnarray}
Thus, one recovers the information encoded when the encoding and decoding $q$-plates have identical topological charges ($q=q'$). This technique is  in principle identical to the modal decomposition of scalar modes with SLMs (see Supplementary Information for further details): a mode is directed onto the SLM, where an inner product of the incident field with a match filter hologram is performed,  and the on-axis intensity is measured by a camera situated after a Fourier lens \cite{flamm2013}. When the input mode matched the filter, a bright on-axis intensity was observed; otherwise a zero on-axis intensity was measured. Thus, the modal content of the state exiting the turbulence plate was efficiently measured.\\

\textbf{Concurrence of entangled qubit pairs.}
In general, the concurrence of an arbitrary qubit state (pure or mixed) can be computed from its density matrix $\rho$ \BP{\cite{Wootters2001}}
\begin{equation}
\mathcal{C}(\rho) = \max\{0,\sqrt{\lambda_1} - \sqrt{\lambda_2} - \sqrt{\lambda_3} - \sqrt{\lambda_4}\} ,
\label{eq: conc_gle}
\end{equation}
where $\lambda_i$ are the eigenvalues in decreasing order of the Hermitian matrix $R = \rho(\sigma_{2}\otimes\sigma_{2})\rho^*(\sigma_{2}\otimes\sigma_{2})$, and $\sigma_{2}$ is the Pauli matrix $\sigma_{2} =  \left[ \begin{smallmatrix} 0 & - i \\ i & 0 \end{smallmatrix} \right]$. The density matrix, $\rho$, of a qubit state, can be expressed in terms of the Pauli matrices \cite{Agnew2011}:
\begin{equation}
\rho = \frac{1}{4} \sigma_0\otimes\sigma_0 + \mathop{\sum_{n,m = 0}}_{n\neq m=0}^{3} \rho_{n,m}\ \sigma_n\otimes\sigma_m\ ,
\label{densitymat}
\end{equation}
where  $\rho_{n,m}$ are complex coefficients, and $\sigma_0$ is the identity matrix. We used Eqs. \ref{eq: conc_gle} and \ref{densitymat} to evaluate experimentally the concurrence of classical and quantum states plotted in Figs. \ref{Fig: Evo Entangl}(a) and \ref{Fig: Evo Entangl}(b)\\

\textbf{Measuring the non-separability of vector vortex modes.}
We applied a tomographic tool to reconstruct the density matrix for the state, so as to analyse the perturbed vector modes. Figure~\ref{Fig:setup}(a) illustrates the experimental setup used. After passing through the turbulence screen, projective measurements were first performed on the polarisation state using a half- and quarter-wave plate, while the OAM DoF was measured using holograms encoded onto the SLM. As SLMs are polarisation sensitive, a linear polariser could not be used to measure the polarisation states, as is commonly performed. Instead, a half-wave plate was inserted before the SLM and rotated to specific orientations to realise a filter for the linear polarisation states: horizontal, vertical, diagonal and anti-diagonal. By inserting a quarter-wave plate, the two circularly polarised components were also filtered, resulting in a total of six polarisation measurements. Similarly, we created six different holograms on the SLM to represent the two pure OAM modes as well as four orientations of the superposition states: $\left|\ell~=~1~\right\rangle~+~\exp(i\theta)\left|\ell~=~-1~\right\rangle$, for $\theta = 0, \pi/2, \pi, 3\pi/2$. A modal decomposition was performed for each polarisation state. 
This tomographic method produces an over-complete set of 36 measurements, which can be used to minimise the $\chi$-square quantity and reconstruct the density matrix $\rho$ \cite{Jack2009}. The concurrence can then be calculated from \BP{Eq. \ref{eq: conc_gle}.}


\textbf{Computing the fidelity between two states.}
In quantum mechanics, the fidelity is a measure of the degree of similitude between an arbitrary state with density matrix $\rho$, and a target state with density matrix $\rho_t$. It is defined as
\begin{equation}
F(\rho,\rho_t) = \left[\text{Tr}\left\{\sqrt{\sqrt{\rho_t}\rho\sqrt{\rho_t}}\right\}\right]^2.
\end{equation}
In our case, we measured the fidelity of a perturbed state $\ket{\Psi_\ell}_{\text{out}}$ with respect to a maximally entangled Bell state $\ket{\Phi}$, for which the density matrix $\rho_t = \ket{\Phi}\bra{\Phi}$, reducing the expression of the fidelity to
\begin{equation}
F(\rho,\rho_t) = \bra{\Phi}\rho\ket{\Phi}.
\end{equation}

\textbf{Modeling the concurrence with respect to the turbulence strength.} The decay of the concurrence of our quantum state can be modelled in terms of the turbulence strength of the channel. Here we consider a turbulence model based on Kolmogorov's theory \cite{kolmogorov1941local}, and use the Strehl ratio (SR) \cite{strehl1894theorie} as our measure of the turbulence strength.  This parameter is defined as the ratio of the on-axis mean irradiance from a point source measured at the plane of a receiver in the presence of turbulence, to that with no turbulence. Assuming weak irradiance fluctuations, we express the turbulence strength as (see Supplementary Information):
\begin{equation}
	{\rm SR} = \frac{1}{1+6.88 (w_0/r_0)^{5/3}}\ ,
	\label{qsfsr}
\end{equation}
\noindent where $w_0$ is the radius of the fundamental (Gaussian) mode and $r_0$ is Fried's parameter given by \cite{fried1966}:
\begin{equation}
	r_0 = 0.185 \left( {\lambda^2\over C_n^2\ z}\right)^{3/5}\ .
	\label{fried}
\end{equation}
In the above expression, the term $C_n^2$ is the refractive index structure parameter, $\lambda$ the wavelength and $z$ the propagation distance. For a single photon (or equivalently one DoF) propagating through turbulence, the concurrence evolves as (see Supplementary Information) \cite{Roux2015}:
\begin{equation}
	{\cal C}(\ket{\Psi_\ell}_{\mathrm{out}}) = \mathrm{\frac{SR}{SR^2-SR+1}}\ .
	\label{concsr}
\end{equation}
\ 

\textbf{Quantum experiment: single photon through turbulence.}
The quantum results in Fig. \ref{Fig: Evo Entangl}(a) were obtained by performing an experiment similar to that in \cite{ibrahim2013}. A 3mm BBO crystal was pumped with a picosecond laser with wavelength of 355nm and an average power of 350mW to produce non-collinear, degenerate entangled photon pairs with type I phase matching via SPDC. Each photon was directed onto a SLM where the conjugate of the modes to be measured and the turbulence was encoded. The modulated beams were then coupled into single mode fibres (SMF), which extract the Gaussian profile from the beams. Avalanche photodiodes were used to register the presence of photon pairs from the SMFs via a coincidence counter.

\section{Acknowledgments}
We express our gratitude to Lorenzo Marrucci for providing us with $q$-plates. B.N. acknowledges financial support from the National Research Foundation of South Africa and the African Laser Centre. C.R.G. acknowledges Claude Leon Foundation. \BP{B.P., C.R.G. and R.I.H. acknowledge support from CONACyT.}


%

\newpage
\section{\label{sec: level5}Supplementary information}

\subsection*{Sorting vector modes using $q$-plates}
Consider an arbitrary vector mode $\ket{\Psi}_{\ell}$  generated by passing a linearly polarised field Gaussian field $A(\vec{r})\hat{i}$ through a $q$-plate, where the position vector $\vec{r} = (r,\phi)$ is expressed in standard polar coordinates, and the unit vector $\hat{i}$ represents the polarisation direction. The Gaussian field is transformed by a $q$-plate, which can be represented by the following Hermitian operator,
\begin{equation}
	\hat{Q}_q = \left[\begin{matrix}
		\cos(2q\phi) & \sin(2q\phi)\\ \sin(2q\phi) & -\cos(2q\phi)
	\end{matrix}\right]\ ,
\end{equation}
where $q$ is the topological charge of the $q$-plate. Subsequently, passing $\ket{\Psi}$ through a second q-plate with topological charge $q'$, and measuring the linear polarisation state results in the following output:
\begin{equation} T = \bra{j}\hat{Q}_{q'}^{\dagger}\hat{Q}_{q}A(\vec{r})\ket{i} = A(\vec{r}) \bra{j}\hat{Q}_{q'}^{\dagger}\hat{Q}_{q}\ket{i} = \braket{\Phi}{\Psi},
	\label{eq: innerprod}
\end{equation}
where $\ket{\Phi}$ is a vector state. Let $\ket{r}$ be a two-dimensional normalised position vector. By projecting Eq. \ref{eq: innerprod} into position space, we obtain
\begin{eqnarray}
	T(r) = \bra{\Phi}\mathbb{I}\ket{\Psi} = \int dr\braket{\Phi}{r}\braket{r}{\Psi} = \Phi^*(r)\Psi(r).
\end{eqnarray}
Using a lens, the field observed in the Fourier plane is given by
\begin{equation}
	T(k) = \int dr\ \Phi^*(r)\Psi(r) \exp(i \vec{k}\cdot \vec{r}). 
\end{equation}
From the orthogonality of OAM modes and polarisation, the on-axis intensity in the Fourier plane will be given by
\begin{equation}
	|T(0)|^2 = |A(\vec{0})|^2\delta_{q,q'}\delta_{i,j}.
\end{equation}
This means that measuring the on-axis intensity will yield a non-zero value if and only if the two $q$-plates have the same topological charge $q$, and the polarisation measured is that of the initial field that generated $\ket{\Psi}$.

\subsection*{Determining the turbulence strength}
\begin{figure}[h]
	\centering
	\includegraphics[width = 0.45\textwidth, height = 0.09\textheight]{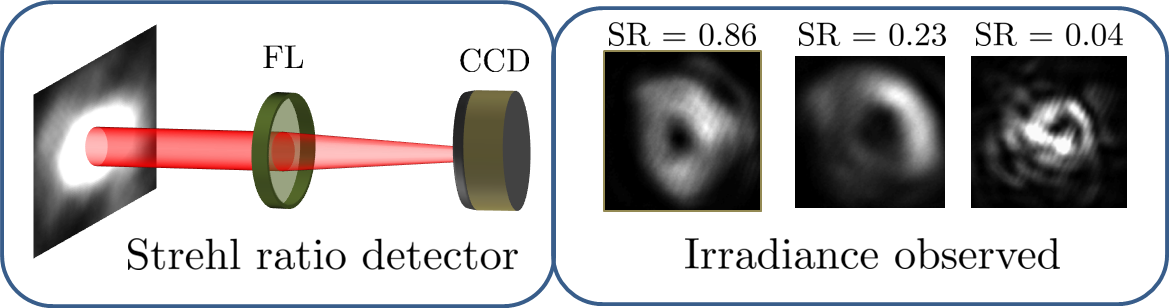}
	\caption{The turbulence strength, given by the Strehl ratio, is measured by the relative drop in a peak intensity between a perturbed and non-perturbed Gaussian beam. }
	\label{Fig: SR setup}
\end{figure}

The strength of a turbulent medium can be characterised by the Strehl ratio \cite{AP}, which is defined as
\begin{equation}
	S_{R} = \frac{I}{I_0},
	\label{eq:Strehl}
\end{equation}
where $I$ and $I_0$ are the on-axis intensities of the aberrated and non-aberrated Gaussian modes, respectively. This is applicable for both the weak and strong turbulence regimes, where 1 represents no turbulence and 0 represents a highly turbulent medium.  Figure~\ref{Fig: SR setup} illustrates the detrimental effects of different turbulence strengths on a vector vortex mode.

\subsection*{Encoding turbulence on an SLM}
\begin{figure}[h]
	\centering
	\includegraphics[width = 0.45\textwidth, height = 0.23\textheight]{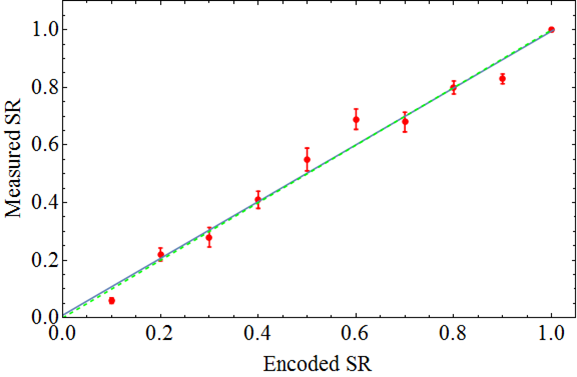}
	\caption{Calibration of turbulence phase screens encoded on a SLM through a comparison of the encoded and measured SR values.}
	\label{Fig: SR calibration}
\end{figure}

The Kolmogorov power spectrum is given by \cite{AP}
\begin{equation}
	\Phi_n(\kappa) = 0.033C_n^2\kappa^{-11/3},
\end{equation}
with $1/L_0\leq\kappa\leq1/l_0$, where $L_0$ and $l_0$ are the inner and outer scales of the turbulence, and define the limits within which the above power spectrum describes an isotropic and homogeneous atmosphere. The turbulence phase screens are generated by Fourier transforming the product of a random function with the power spectrum above. Using a SLM, we digitally generated turbulence phase screens and obtained the calibration curve illustrated in Fig. \ref{Fig: SR calibration}

\subsection*{Determining the concurrence in terms of the Strehl ratio}

The effect of a turbulent medium on the Strehl ratio, assuming weak irradiance fluctuations, is given by \cite{AP}
\begin{equation}
	{\rm SR} \approx \frac{1}{1+(D/r_0)^{5/3}}\ ,
	\label{Eq: SR1}
\end{equation}
where SR is the Strehl ratio, $D$ is the diameter of the receiving aperture, and $r_0$ is the Fried parameter \cite{fried}, given by 
\begin{equation}
	r_0 = 0.185 \left( {\lambda^2\over C_n^2 z}\right)^{3/5}\ .
	\label{Eq: fried}
\end{equation}
Although Eq.~(\ref{Eq: SR1}) is valid for weak irradiance fluctuations, it has not been derived for a single phase screen scenario, which is the case for the current experimental setup. One can compute the Strehl ratio for a single phase screen, using the quadratic structure function approximation \cite{leader}. The resulting expression
\begin{equation}
	{\rm SR} = \frac{1}{1+6.88 (w_0/r_0)^2}\ ,
	\label{Eq: SR2}
\end{equation}
is similar to Eq.~(\ref{Eq: SR1}). Here $w_0$ is the beam radius of the input beam. We will assume that, without the quadratic structure function approximation, the relationship is of the form 
\begin{equation}
	{\rm SR} = \frac{1}{1+6.88 (w_0/r_0)^{5/3}}\ .
	\label{Eq: SR3}
\end{equation}

The concurrence of a photon pair that has an initial entangled state (Bell state)
\begin{equation}
	\ket{\Psi^{+}} = \frac{1}{\sqrt{2}}\left( \ket{\ell}\ket{{-\ell}} + \ket{{-\ell}}\ket{\ell} \right)\ ,
	\label{bell}
\end{equation}
and where only one photon propagates through single phase screen turbulence, evolves according to
\begin{equation}
	{\cal C}\left(\ket{\Psi^{+}}\right) = \frac{X+1}{X^2+X+1}\ ,
	\label{conc}
\end{equation}
where 
\begin{equation}
	X = 6.88 (w_0/r_0)^{5/3}\ .
	\label{xdef}
\end{equation}
Using Eq.~(\ref{Eq: SR3}), we can express $X$ in terms of the Strehl ratio  
\begin{equation}
	X =\frac{1}{{\rm SR}} - 1\ ,
	\label{xsr}
\end{equation}
so that Eq.~(\ref{conc}) becomes
\begin{equation}
	{\cal C}\left(\ket{\Psi^{+}}\right) = \frac{{\rm SR}}{{\rm SR}^2-{\rm SR}+1}\ .
\end{equation}
\tk{\subsection*{Error correction for turbulence}
A particular realization of the turbulent atmosphere acts  on the spatial modes due to modal cross-talk and detection of a subspace as a filter with a single rank-two Kraus operator M (Eq. \ref{Eq: channelmat}), which can be expressed in polar decomposition as 
\begin{equation} M =U |M|= U (\lambda_0 \ket{0}\bra{0 } + \lambda_1\ket{1}\bra{1 }),  \end{equation} 
where $U$ is unitary and $\lambda_0 \ket{0}\bra{0 } + \lambda_1\ket{1}\bra{1 }$ is the spectral decomposition of the positive operator $|M|= \sqrt{M^\dagger M}$.
The action of the filter $M$ can  be compensated by a second `conjugate' filter $\tilde{M}$ with $\tilde{M}M\propto \mathbb{1}$ given by
\begin{equation} \tilde{M}=  (\lambda_1 \ket{0}\bra{0 } + \lambda_0\ket{1}\bra{1 }) U^\dagger,   \end{equation} 
which can be physically implemented.  In the example of error correction discussed in the paper, the action of the channel $M$ was compensated by processing the measurement data with $M^{-1}$. }   
\subsection*{Effects of OAM crosstalk on the concurrence}
For a hybrid OAM-polarisation qubit state $\ket{\Psi}_{\ell}~=~\alpha\ket{\ell}\ket{R} + \gamma\ket{\ell}\ket{L} + \beta\ket{-\ell}\ket{L} + \tau\ket{-\ell}\ket{R}$, the concurrence is computed as follows \cite{Wootters2001_2}
\begin{equation}
	\mathcal{C}(\ket{\Psi}_{\ell}) = 2|\alpha\beta - \gamma\tau|.
\end{equation}
For a vector vortex mode defined as in Eq. \ref{eq: vector beam_classical}, reduces to $\mathcal{C}(\ket{\Psi}) = 2|\alpha\beta|$.

Recall the expression derived for the concurrence of the input and output states:
\TK{\begin{equation}
	\mathcal{C}(\ket{\Psi_\ell}_{\text{out}}) = |p_0^2 - p_{2\ell}p_{-2\ell}|~\mathcal{C}(\ket{\Psi_\ell}_{\text{in}}),
	\label{eq:coutcin1}
\end{equation}
 }
\tk{where we omitted a normalization constant for the sake of simplicity}. We can extend our analysis by imposing conditions on $p_{\ell}$. We want a symmetric distribution with its maximum $p_\ell=1$ centered at $\ell=0$. For the sake of the argument, we will assume a Gaussian-like discrete function for $p_\ell$
\begin{equation}
	p_\ell = \exp(-{\ell}^2/2\Delta^2),
\end{equation}
where $\ell$ is the OAM index and $\Delta$ is the width of the distribution, which depends on turbulence. We can rewrite Eq. \ref{eq:coutcin1} as:
\begin{equation}\label{eq:coutcin2}
	\mathcal{C}(\ket{\Psi_{\text{out}}}) = |1 - \exp(-\ell^2/\Delta^2)|\ \mathcal{C}(\ket{\Psi_{\text{in}}}).
\end{equation}

\begin{figure}[h!]
	\centering
	\includegraphics[width=80mm,height=96mm]{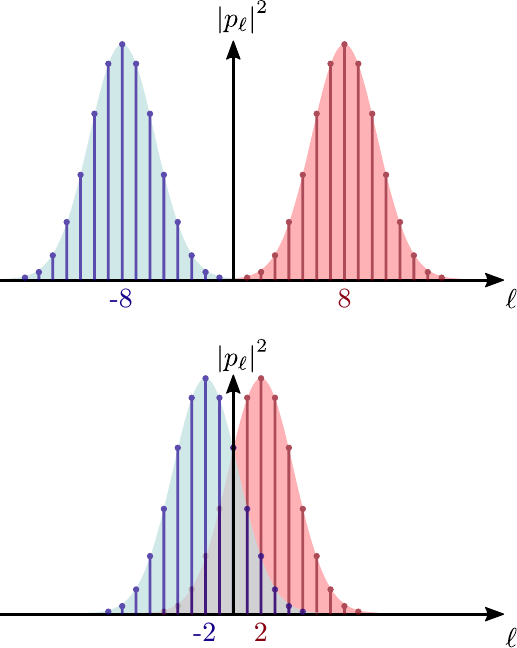}
	\caption{Schematic representation of OAM crosstalk. Increasing the separation of the modes or decreasing the width of each Gaussian curve (decreasing turbulence) reduces the crosstalk.}
	\label{fig:gaussianturb}
\end{figure}

If $\ell = 0$, then the concurrence $C(\ket{\Psi_\ell}_{\text{out}}) = 0$, which is explained by the fact that the input beam is not a vector beam. In the case of $\ell\to\infty$, the concurrence of the output state will be equal to that of the input state: $C(\ket{\Psi_\ell}_{\text{out}}) = C(\ket{\Psi_\ell}_{\text{in}})$. This is because the OAM modes are so far apart that the crosstalk resulting from the turbulence will not affect the measured OAM modes, as illustrated in Fig. \ref{fig:gaussianturb}. If $\Delta = 0$, then $C(\ket{\Psi_\ell}_{\text{out}}) = C(\ket{\Psi_\ell}_{\text{in}})$, as this implies that the initial state is not perturbed (no turbulence).

\end{document}